\documentclass[
reprint,
superscriptaddress,
amsmath,amssymb,
aps,
prapplied,
groupedaddress
]{revtex4-2}

\usepackage{graphicx}
\graphicspath{{.}} 
\usepackage{dcolumn}
\usepackage{bm}
\usepackage{physics}
\usepackage{braket}
\usepackage{placeins}
\usepackage{upgreek}
\usepackage{dsfont}

\usepackage[colorlinks=true]{hyperref}

\begin{document}

\let\oldaddcontentsline\addcontentsline
\renewcommand{\addcontentsline}[3]{}

\title{In-situ characterization of qubit drive-phase distortions}

\begin{abstract}
    Reducing errors in quantum gates is critical to the development of quantum computers.
    To do so, any distortions in the control signals should be identified, however, conventional tools are not always applicable when part of the system is under high vacuum, cryogenic, or microscopic.
    Here, we demonstrate a method to detect and compensate for amplitude-dependent phase changes, using the qubit itself as a probe.
    The technique is implemented using a microwave-driven trapped ion qubit, where correcting phase distortions leads to a three-fold improvement in the error of single-qubit gates implemented with pulses of different amplitudes, to attain state-of-the-art performance benchmarked at $1.6(4)\times 10^{-6}$ error per Clifford gate.
\end{abstract}

\author{M.\,F.\,Gely}
\thanks{These two authors contributed equally}

\author{J.\,M.\,A.\,Litarowicz}
\thanks{These two authors contributed equally}

\author{A.\,D.\,Leu}

\author{D.\,M.\,Lucas}

\affiliation{Clarendon Laboratory, Department of Physics, University of Oxford, Parks Road, Oxford OX1 3PU, U.K.}

\date{\today}

\maketitle


\begin{figure}[t]
    \centering
    \includegraphics[width=0.45\textwidth]{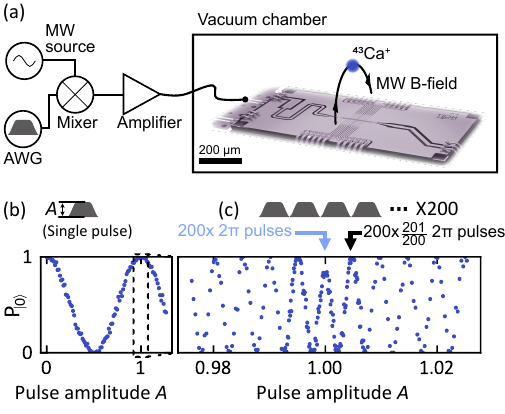}
    \caption{
    \textbf{Non-linear drive chain and Rabi-oscillation distortion.}
    (a) Overview of the experimental setup showing a simplified schematic of the MW drive chain delivering pulses to a $^\text{43}$Ca$^+$ trapped-ion qubit.
    The potential nonlinearity of the MW components needs to be identified and compensated for.
    Ion height and MW field not to scale.
    (b) Delivering a single pulse of fixed duration, but varied amplitude $A$, enables the observation of Rabi-oscillations.
    Here $P_{\ket{0}}$ corresponds to the probability of measuring the initial state $\ket{0}$ after the pulse.
    The amplitude is normalized such that $A=1$ corresponds to a pulse driving a $2\pi$ rotation of the qubit.
    (c) When $N$ pulses are delivered, the initial qubit state is in principle recovered when the pulse amplitude is an integer multiple of $1/N$.
    Here this is verified for $N=200$, and surprisingly is not the case for $A=1$.
    This exceptional point corresponds to the case where $N\times \ 2\pi$ pulses are nominally delivered.
    As explained in Fig.~\ref{fig:fig2}, this is a signature of a phase-amplitude relation, stemming from the nonlinearity of MW components.
    }
    \label{fig:fig1}
    \end{figure}

\section{Introduction}

High-fidelity control of quantum systems is crucial to quantum computing, as it can drastically minimize the overhead associated with error correction~\cite{knill2010}.
In many platforms, quantum gates are driven by laser, radio-frequency or microwave pulses, where low error rates depend on the accurate control of signal amplitude and phase.
Distortions induced by the chain of components generating and delivering this drive signal should therefore be identified and corrected.
A common compensation technique is to use ``pre-distortion'' such that distortions produce the desired pulse~\cite{Hincks2015}. 
This is used in NMR~\cite{tabuchi2010}, 
EPR~\cite{Spindler2017}, 
Rydberg atoms~\cite{singh2023compensating}, 
Bose-Einstein condensates~\cite{jager2013},
and superconducting circuits~\cite{Gustavsson2013,Jerger2019, lazifmmode2023calibration}.
%
In trapped ions~\cite{bruzewicz2019trapped}, pre-distortion has been used to compensate for thermal transients in microwave circuitry~\cite{harty2016}, or for compensating filtering in fast shuttling operations~\cite{bowler2013}.
Successful pre-distortion relies however on an accurate model of the drive chain, part of which is, for trapped ions, under high vacuum, microscopic and possibly cooled cryogenically~\cite{poitzsch1996cryogenic}.
Using conventional tools such as network analyzers is therefore not always possible.

Here, we demonstrate how a trapped ion qubit can be used as a sensor to characterize the distortion of gate pulses~\cite{Gustavsson2013,Jerger2019, lazifmmode2023calibration,kristen2020amplitude}.
This complements other techniques relying on atomic systems such as Bose-Einstein condensates~\cite{Ockeloen2013} or ultracold atoms~\cite{Bohi2010} to image near-field microwaves.
Specifically, we present a scheme which coherently amplifies the amplitude-dependent phase change of a microwave field.
Pre-distortion is shown to compensate for this drive chain imperfection, leading to a three-fold reduction in single-qubit gate error when gates are implemented with pulses of different amplitudes, to attain state of the art performance benchmarked at $1.6(4)\times 10^{-6}$ error per Clifford gate.

\begin{figure}[t]
\centering
\includegraphics[width=0.45\textwidth]{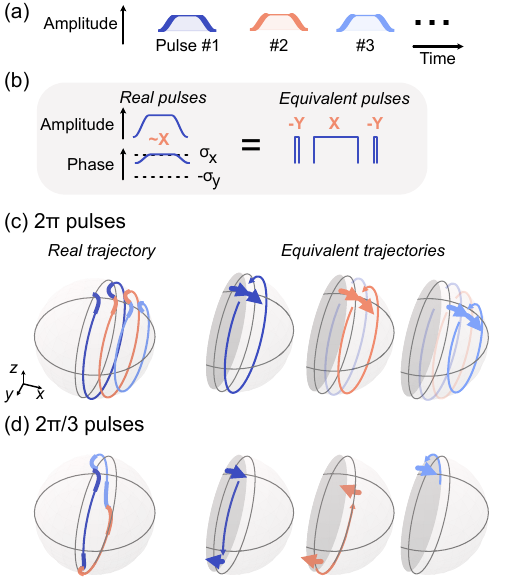}
\caption{
\textbf{Coherent amplification of phase variations.}
(a) Schematic of the train of pulses applied to the qubit.
These are intended to produce $x$-rotations.
(b) If the MW phase depends on amplitude (``real pulses''), the resulting deviations in phase during ramp up/down can be described by a small $y$-rotation (``equivalent pulses'').
The trajectory driven on the Bloch sphere by three successive pulses is shown in (c) and (d) for the real and equivalent pulses.
The ramping or resulting $y$-rotations are shown in bold or with a thicker arrow.
(c) In the case of a train of $2\pi$ pulses, since the qubit returns to the same part of the Bloch sphere for ramp-up/down of the pulse amplitude, the small $y$-rotations add up coherently, producing the exceptional $A=1$ point in Fig.~\ref{fig:fig1}(c).
(d) Otherwise, for example here with $2\pi/3$ pulses, ramp-up/down occurs on varying parts of the Bloch sphere, such that the $y$-rotations cancel out, ensuring a return to the initial qubit state.
To make the effect visible for just 3 pulses, ramp durations and phase deviations are exaggerated with respect to the experiment.
}
\label{fig:fig2}
\end{figure}

\begin{figure*}[t]
\centering
\includegraphics[width=0.9\textwidth]{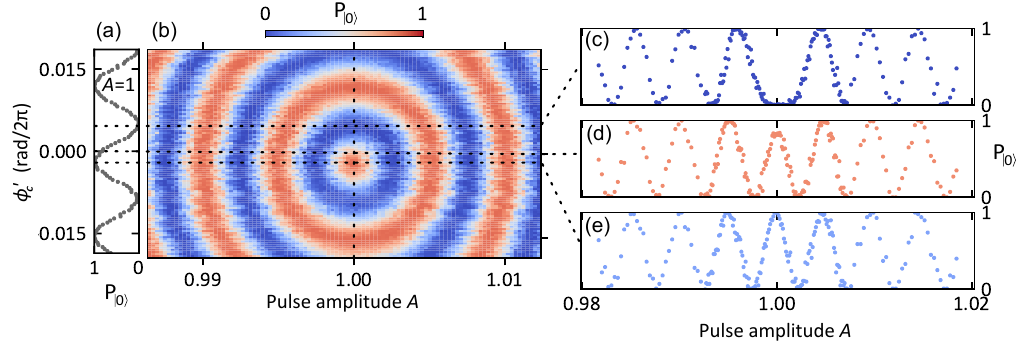}
\caption{
\textbf{MW phase pre-distortion.}
Rabi-oscillations are measured as in Fig.~\ref{fig:fig1}(c) for a train of $N=200$ pulses.
(a) For $A=1$, we vary the proportionality factor between phase and amplitude $\phi_c'$, implemented in our AWG, whilst monitoring $P_{\ket{0}}$.
Compensation of the effect is obtained for $\phi_c'=-2\pi\times 1.8\times 10^{-3}$ radians.
(b) $P_{\ket{0}}$ as a function of both $\phi_c'$ and pulse amplitude $A$, where selected transects show that the effect can be accentuated (c, $\phi_c'=2\pi\times4.9\times10^{-3}$) with respect to the case where the phase is not intentionally varied (d, $\phi_c'=0$), or compensated for (e).
}
\label{fig:fig3}
\end{figure*}

\section{Apparatus and experimental observations}

Experiments are carried out using a trapped ion qubit driven by near-field microwaves~\cite{ospelkaus2008,Warring2013Phase,harty2014,harty2016,zarantonello2019,srinivas2021,leu2023}.
Our qubit is defined by the hyperfine levels $\ket{0}=|F=3, M_{F}=1 \rangle$ and $\ket{1}=|F=4, M_{F}=1 \rangle$ of the ground state manifold 4S$_{1/2}$ of $^{43}$Ca$^{+}$, which form a clock transition at our static magnetic field strength of 28.8 mT.
The ion is trapped $40\ \upmu$m above a surface ``chip'' trap, which features a microwave (MW) electrode used to drive the magnetic dipole moment of the qubit.
Amplitude-shaped $\sim100$ MHz pulses are generated using an arbitrary waveform generator (AWG), then up-converted using a $3$ GHz source and an IQ mixer to be brought into on resonance with the qubit frequency at 3.1 GHz.
After amplification to $\sim$ 400 mW, the microwaves are delivered into the vacuum chamber to the surface trap where they drive the qubit.
A simplified schematic of the experimental setup is shown in Fig.~\ref{fig:fig1}(a), with more detail in Sec.~\ref{sec:supporting_measurements}.
Details on the trap and qubit state preparation and readout can be found in Ref~\cite{weber2022cryogenic}.

These MW pulses, resonant with the qubit transition, will drive Rabi-oscillations, as shown in Fig.~\ref{fig:fig1}(b).
Given our control system's 1 ns timing resolution and our pulse duration of $1.2\ \upmu$s, varying the pulse amplitude at the AWG (with 15-bit resolution) offers a more precise method of tuning the pulse area than varying the pulse duration ($3\times10^{-5}$ precision rather than $8\times10^{-4}$).
We therefore vary pulse amplitude rather than the more usual pulse time in these measurements, but scanning pulse duration would produce a similar effect in Figs.~\ref{fig:fig1}(b,c).
For a fixed pulse shape and duration, we define a pulse amplitude scaling factor $A$ such that $A=1$ produces a $2\pi$ rotation of the qubit.
When $A$ is swept from 0 to 1, the probability $P_{\ket{0}}$ of recovering the initial qubit state $\ket{0}$ will undergo a single oscillation as shown in Fig.~\ref{fig:fig1}(a).
When driving the qubit with a train of $N$ (identical) pulses rather than one pulse, $P_{\ket{0}}$ should undergo $N$ oscillations rather than one in the same amplitude span.
Indeed, at a pulse amplitude $A = n/N$, if $n$ is an integer, the cumulative effect of the pulse train should be to induce $n$ full rotations of the qubit state around the Bloch sphere.
The resulting recovery to the initial state is observed in the measurement of Fig.~\ref{fig:fig1}(c), except for the case $A=1$, nominally corresponding to a train of $2\pi$ rotations.

\section{Theoretical description of the phenomenon}

To explain this phenomenon, we consider a change in MW phase $\phi(t)$ during the pulse, proportional to the pulse amplitude $a(t)$.
We normalize $a$ such that $a(t)=A$ during the ``plateau'' of the pulse.
The phase variation is assumed to be characterized by the phase-amplitude proportionality factor $\phi' = \partial\phi/\partial a$, such that we have $\phi(t) = \phi' a(t) +  \phi_{0}$.
Such a phase-amplitude relation would come into play whilst ramping up/down the pulse, which is done in this experiment over a duration of $t_\text{ramp}=$200 ns per ramp (400 ns total ramping time over a pulse).
This consideration is motivated by the number of potentially non-linear elements in the MW drive chain (AWG, mixer, amplifier).
Without loss of generality, we set $\phi(t)=0$ during the ``plateau'' of the pulse such that the small phase variation perturbs the pulse -- during the ramping -- away from the ideal $x$-rotation.
The resulting physics can be understood by constructing an equivalent unitary to that driven by the pulse train, where phase deviations manifest as small $y$-rotations at the beginning and end of each pulse train.
This is schematically shown in Fig.~\ref{fig:fig2}(b) and formally derived in Sec.~\ref{sec:ZYXYZ_decomposition}.
These small $y$-rotations will nearly always be inconsequential as the dominant $x$-rotation will dynamically decouple the qubit from non-commuting $y$-rotations.
However, this is not the case at the exceptional $A=1$ point in Fig.~\ref{fig:fig1}(c), where the $x$-rotation is an identity operation, which then commutes with $y$-rotations.
Indeed, at $A=1$, each pulse is approximately a $2\pi$ rotation around the $x$-axis, and ramping up/down the pulse always occurs when the qubit state is in the same area of the Bloch sphere.
The $y$-rotations will then coherently add up such that, after $N=200$ pulses, the total $y$-rotation produces a noticeable decrease in $P_{\ket{0}}$.
This is schematically shown in Fig.~\ref{fig:fig2}(c), with inflated ramping durations and phase variations such that the effect is noticeable after a few pulses.
When pulses do not produce a $2\pi$ rotation, ramping up/down will occur at varying areas of the Bloch sphere as the train of pulses progresses, and the $y$-rotations tend to average out.
This is illustrated in the case of $2\pi/3$ pulses in Fig.~\ref{fig:fig2}(d).
As derived in Sec.~\ref{sec:perturbation_theory}, in the former case phase variations have a $\sim(N\phi')^2$ impact on $P_{\ket{0}}$, whereas in the latter the effect is much smaller $\sim(N\phi')^4$.

\section{Experimental verification}

To verify this interpretation, we add a compensation phase proportional to the pulse amplitude $\phi_c(t) = a(t)\phi_c'$ in the pulses generated by our AWG.
This ``pre-distorted'' MW pulse then has a phase given by
\begin{equation}
    \phi(t) = \phi_n(t)+\phi_c(t) = a(t)(\phi_n'+\phi_c') + \phi_{0}
\end{equation}
where $\phi_n$ designates the phase variation native to the drive chain.
Whilst varying the parameter $\phi_c'$, we measure the distorted Rabi-oscillations to show that full contrast at $A=1$ can be recovered for $\phi_c'= -2\pi\times 1.8\times10^{-3}$ radians (Fig.~\ref{fig:fig3}).
Here we are again using a train of 200 pulses just as in Fig.~\ref{fig:fig1}(c).
This indicates that the phase does indeed change (linearly) with MW amplitude. 
A control experiment providing further confirmation of this fact is presented in Sec.~\ref{sec:supporting_measurements}.

Scanning both pulse amplitude $A$ and phase compensation $\phi_c'$ whilst monitoring $P_{\ket{0}}$ reveals a symmetry in their influence (Fig.~\ref{fig:fig3}(b)).
Indeed, whilst deviations of the pulse amplitude away from $A=1$ increases the amount of net rotation around the $x$-axis, increased phase variation during pulse ramping increases the accumulated $y$-rotation over the train of pulses.
While a horizontal line-cut of Fig.~\ref{fig:fig3} (Fig.~\ref{fig:fig3}(e)) describes Rabi-oscillations induced by $x$-rotations, a vertical line-cut (Fig.~\ref{fig:fig3}(a)) describes Rabi-oscillations induced by $y$-rotations.
This is demonstrated more formally in Sec.~\ref{sec:perturbation_theory} using perturbation theory, showing that for $|A-1|,|\phi(t)|\ll 1$, a train of $N$ pulses drives the unitary evolution
\begin{equation}
    \hat U = \text{exp}\left(iN\Omega_x\hat\sigma_x+iN\Omega_y\hat\sigma_y\right)
\end{equation}
with $\Omega_x\propto(A-1)$ and $\Omega_y\propto\phi'$.
This unitary results in the ring pattern observed in the measurement of Fig.~\ref{fig:fig3}(b):
\begin{equation}
    P_{\ket{0}} = \frac12\left(1+\cos\left(N\sqrt{\Omega_x^2+\Omega_y^2}\right)\right)\ .
\end{equation}

\begin{figure}[t]
    \centering
    \includegraphics[width=0.45\textwidth]{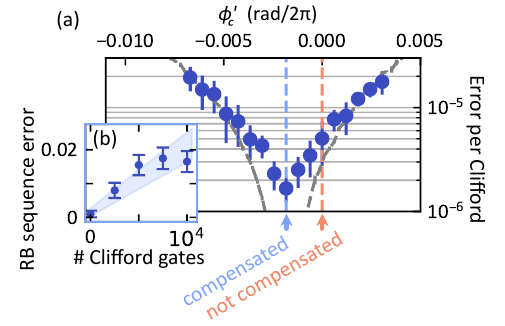}
    \caption{
    \textbf{Application to high-fidelity quantum gates.}
    (a) Dots are single-qubit gate errors measured using randomized benchmarking (RB) as a function of phase compensation.
    The error is reduced from $5(2)\times 10^{-6}$ to $1.6(4)\times 10^{-6}$ when using the phase compensation identified in Fig.\ref{fig:fig3}(a).
    The error scaling at higher phase compensation values is consistent with simulated RB data (gray dashed lines).
    Lower errors are limited by other physical mechanisms~\cite{leu2023}.
    (b)
    RB sequence error as a function of the number of Clifford gates per sequence.
    Corresponds to lowest error where the phase variations are compensated for.
    The gate error is on par with state-of-the art errors across all quantum computing platforms~\cite{harty2014,leu2023}.
    }
    \label{fig:fig4}
    \end{figure}

\section{Application to single-qubit gates}

Whilst the phase change over a non-compensated pulse is small, it can be a leading source of error in the high-fidelity logic operations on trapped-ion qubits.
Scanning the compensation parameter whilst monitoring $P_{\ket{0}}$ (Fig.~\ref{fig:fig3}(a)) constitutes a rapid calibration procedure sufficient to mitigate phase variations.
To demonstrate this, we scan the phase compensation parameter $\phi_c'$ whilst monitoring the single-qubit Clifford gate error with randomized benchmarking~\cite{Knill2008RBM} (Fig.~\ref{fig:fig4}(a)).
The error is reduced from $5(2)\times 10^{-6}$ to $1.6(4)\times 10^{-6}$ when using the phase compensation identified in Fig.~\ref{fig:fig3}(a).
In this case, identifying and compensating for this small variation in phase brings single-qubit gate performance to state of the art error rates~\cite{leu2023,harty2014}.
The sensitivity the technique, considering the presence of decoherence, drifts or state-preparation and measurement errors, is discussed in Sec.~\ref{sec:sensitivity}.
This calibration routine could be extended to more complex phase-amplitude relations $\phi(t)=\phi_0+\phi'a(t)+\phi''a(t)^2+\phi'''a(t)^3+...$ by maximizing $P_{\ket{0}}$ as in Fig.~\ref{fig:fig3}(a) as a function of $\phi'$ but also higher order phase compensation parameters $\phi''$, $\phi'''$ .

We note that this problem is not present in previously reported error budgets~\cite{leu2023} due to the different decomposition of Clifford gates into native MW pulses.
Here, we decompose Clifford gates into $\pi$ and $\pi/2$ pulses of the same duration, and the bulk of the error arises due to the $\times2$ variation in pulse amplitude, which produces a change in the relative phases of $\pi$ and $\pi/2$ pulses.
This error can thus also be mitigated by decomposing Clifford gates into only $\pi/2$ pulses, at the expense of increasing the average number of pulses per Clifford from $\approx$1 to $\approx$2.2.
Since the pulse time is dominated by the inter-pulse delay imposed by our control system, this increases the effective gate time and the impact of decoherence.
This option was utilized in Ref.~\cite{leu2023}, leading to similar gate errors.
The residual error from phase variation during pulse ramping is then negligeable.
Alternatively, $\pi$ pulses can be implemented with twice the duration of $\pi/2$ pulses, with pulse area fine-tuned through pulse amplitude, minimizing the impact of phase variations, and the average duration of a Clifford gate. 

A calibration of phase variations is also useful in more complex composite pulse schemes.
In particular when we want to leverage the fine pulse amplitude control from pulse to pulse -- for example in the ion addressing scheme of Ref.~\cite{leu2023}.
In the addressing experiment, the combination of weak pulses (at least half the amplitude used in this work), and the dominance of drift-induced errors made the impact of phase variations negligeable.
The average error over 100 different addressing pulse sequences is theoretically $9.0(2)\times 10^{-7}$, much smaller than the average error measured at $3.4\times 10^{-5}$.
However, by simply doubling the amplitude and  $\phi_n'$, the error would have been considerable at $1.5(5)\times 10^{-5}$ per gate.
By measuring a scattering parameter ($S_\text{21}$) of the nonlinear components of our MW drive chain, we attempt to understand the native phase variation characterized by $\phi_n'= 2\pi\times 1.8\times10^{-3}$ rad.
Unfortunately, such measurements are limited to the phase precision of our vector network analyzer  $\sim 2\pi\times 10^{-4}$ rad.
The only component showing measurable phase variation is the amplifier with $\phi'= 2\pi\times 0.8\times10^{-3}$, see measurement details in Sec.~\ref{sec:supporting_measurements}.
In this experimental context, the Rabi-oscillation distortion is thus a much more sensitive method for diagnostic and compensation.

Finally, we note that the measurement scheme of Fig.~\ref{fig:fig1}(c) also enables the characterization of other drive chain distortions.
If the Rabi-frequency scales linearly with the MW amplitude programmed in the AWG, then the period of oscillations in Fig.~\ref{fig:fig1}(c) should remain constant as the amplitude increases.
In practice, we find that this is not the case, and this measurement constitutes a rapid calibration of this field-amplitude non-linearity which could be useful for e.g. the addressing scheme of Ref.~\cite{leu2023}, which relies on pulses of varying amplitudes.

\section{Conclusion}

In conclusion, we have presented a scheme to rapidly identify, and to compensate for, phase-amplitude relations in qubit drives.
These phase variations are shown to be a dominant error in our implementation of microwave-driven quantum logic.
Pre-distortion of the phase evolution during the pulse ramping is shown to be an effective mitigation technique.
The resulting error rate, measured through randomized benchmarking ($1.6(4)\times 10^{-6}$ per Clifford gate), is consistent with the state of the art across all quantum computing platforms~\cite{harty2014,leu2023}.
The scheme is applicable to any qubit system driven by an amplitude-shaped pulse.
Since the qubit itself is used to probe the field, it is particularly suited to systems partially inaccessible by more conventional sensors; systems under vacuum, cryogenic, or micro-/nano-meter in size.
For example, the technique could readily be applied to spin qubits, where single-qubit gates are also implemented using gigahertz-frequency shaped pulses, resonantly driving the qubit transition~\cite{Takeda2016}.

\section*{Acknowledgments}
This work was supported by the U.S. Army Research Office (ref. W911NF-18-1-0340) and the U.K. EPSRC Quantum Computing and Simulation Hub.
M.F.G. acknowledges support from the Netherlands Organization for Scientific Research (NWO) through a Rubicon Grant.
A.D.L. acknowledges support from Oxford Ionics Ltd.

\bibliography{library}

\begin{thebibliography}{25}%
\makeatletter
\providecommand \@ifxundefined [1]{%
 \@ifx{#1\undefined}
}%
\providecommand \@ifnum [1]{%
 \ifnum #1\expandafter \@firstoftwo
 \else \expandafter \@secondoftwo
 \fi
}%
\providecommand \@ifx [1]{%
 \ifx #1\expandafter \@firstoftwo
 \else \expandafter \@secondoftwo
 \fi
}%
\providecommand \natexlab [1]{#1}%
\providecommand \enquote  [1]{``#1''}%
\providecommand \bibnamefont  [1]{#1}%
\providecommand \bibfnamefont [1]{#1}%
\providecommand \citenamefont [1]{#1}%
\providecommand \href@noop [0]{\@secondoftwo}%
\providecommand \href [0]{\begingroup \@sanitize@url \@href}%
\providecommand \@href[1]{\@@startlink{#1}\@@href}%
\providecommand \@@href[1]{\endgroup#1\@@endlink}%
\providecommand \@sanitize@url [0]{\catcode `\\12\catcode `\$12\catcode
  `\&12\catcode `\#12\catcode `\^12\catcode `\_12\catcode `\%12\relax}%
\providecommand \@@startlink[1]{}%
\providecommand \@@endlink[0]{}%
\providecommand \url  [0]{\begingroup\@sanitize@url \@url }%
\providecommand \@url [1]{\endgroup\@href {#1}{\urlprefix }}%
\providecommand \urlprefix  [0]{URL }%
\providecommand \Eprint [0]{\href }%
\providecommand \doibase [0]{https://doi.org/}%
\providecommand \selectlanguage [0]{\@gobble}%
\providecommand \bibinfo  [0]{\@secondoftwo}%
\providecommand \bibfield  [0]{\@secondoftwo}%
\providecommand \translation [1]{[#1]}%
\providecommand \BibitemOpen [0]{}%
\providecommand \bibitemStop [0]{}%
\providecommand \bibitemNoStop [0]{.\EOS\space}%
\providecommand \EOS [0]{\spacefactor3000\relax}%
\providecommand \BibitemShut  [1]{\csname bibitem#1\endcsname}%
\let\auto@bib@innerbib\@empty
\bibitem [{\citenamefont {Knill}(2010)}]{knill2010}%
  \BibitemOpen
  \bibfield  {author} {\bibinfo {author} {\bibfnamefont {E.}~\bibnamefont
  {Knill}},\ }\bibfield  {title} {\bibinfo {title} {Quantum computing},\ }\href
  {https://doi.org/10.1038/463441a} {\bibfield  {journal} {\bibinfo  {journal}
  {Nature}\ }\textbf {\bibinfo {volume} {463}},\ \bibinfo {pages} {441}
  (\bibinfo {year} {2010})}\BibitemShut {NoStop}%
\bibitem [{\citenamefont {Hincks}\ \emph {et~al.}(2015)\citenamefont {Hincks},
  \citenamefont {Granade}, \citenamefont {Borneman},\ and\ \citenamefont
  {Cory}}]{Hincks2015}%
  \BibitemOpen
  \bibfield  {author} {\bibinfo {author} {\bibfnamefont {I.~N.}\ \bibnamefont
  {Hincks}}, \bibinfo {author} {\bibfnamefont {C.~E.}\ \bibnamefont {Granade}},
  \bibinfo {author} {\bibfnamefont {T.~W.}\ \bibnamefont {Borneman}},\ and\
  \bibinfo {author} {\bibfnamefont {D.~G.}\ \bibnamefont {Cory}},\ }\bibfield
  {title} {\bibinfo {title} {Controlling quantum devices with nonlinear
  hardware},\ }\href {https://doi.org/10.1103/PhysRevApplied.4.024012}
  {\bibfield  {journal} {\bibinfo  {journal} {Phys. Rev. Appl.}\ }\textbf
  {\bibinfo {volume} {4}},\ \bibinfo {pages} {024012} (\bibinfo {year}
  {2015})}\BibitemShut {NoStop}%
\bibitem [{\citenamefont {Tabuchi}\ \emph {et~al.}(2010)\citenamefont
  {Tabuchi}, \citenamefont {Negoro}, \citenamefont {Takeda},\ and\
  \citenamefont {Kitagawa}}]{tabuchi2010}%
  \BibitemOpen
  \bibfield  {author} {\bibinfo {author} {\bibfnamefont {Y.}~\bibnamefont
  {Tabuchi}}, \bibinfo {author} {\bibfnamefont {M.}~\bibnamefont {Negoro}},
  \bibinfo {author} {\bibfnamefont {K.}~\bibnamefont {Takeda}},\ and\ \bibinfo
  {author} {\bibfnamefont {M.}~\bibnamefont {Kitagawa}},\ }\bibfield  {title}
  {\bibinfo {title} {Total compensation of pulse transients inside a
  resonator},\ }\href
  {https://doi.org/https://doi.org/10.1016/j.jmr.2010.03.014} {\bibfield
  {journal} {\bibinfo  {journal} {Journal of Magnetic Resonance}\ }\textbf
  {\bibinfo {volume} {204}},\ \bibinfo {pages} {327} (\bibinfo {year}
  {2010})}\BibitemShut {NoStop}%
\bibitem [{\citenamefont {Spindler}\ \emph {et~al.}(2017)\citenamefont
  {Spindler}, \citenamefont {Schöps}, \citenamefont {Kallies}, \citenamefont
  {Glaser},\ and\ \citenamefont {Prisner}}]{Spindler2017}%
  \BibitemOpen
  \bibfield  {author} {\bibinfo {author} {\bibfnamefont {P.~E.}\ \bibnamefont
  {Spindler}}, \bibinfo {author} {\bibfnamefont {P.}~\bibnamefont {Schöps}},
  \bibinfo {author} {\bibfnamefont {W.}~\bibnamefont {Kallies}}, \bibinfo
  {author} {\bibfnamefont {S.~J.}\ \bibnamefont {Glaser}},\ and\ \bibinfo
  {author} {\bibfnamefont {T.~F.}\ \bibnamefont {Prisner}},\ }\bibfield
  {title} {\bibinfo {title} {Perspectives of shaped pulses for epr
  spectroscopy},\ }\href
  {https://doi.org/https://doi.org/10.1016/j.jmr.2017.02.023} {\bibfield
  {journal} {\bibinfo  {journal} {Journal of Magnetic Resonance}\ }\textbf
  {\bibinfo {volume} {280}},\ \bibinfo {pages} {30} (\bibinfo {year} {2017})},\
  \bibinfo {note} {special Issue on Methodological advances in EPR spectroscopy
  and imaging}\BibitemShut {NoStop}%
\bibitem [{\citenamefont {Singh}\ \emph {et~al.}(2023)\citenamefont {Singh},
  \citenamefont {Zeier}, \citenamefont {Calarco},\ and\ \citenamefont
  {Motzoi}}]{singh2023compensating}%
  \BibitemOpen
  \bibfield  {author} {\bibinfo {author} {\bibfnamefont {J.}~\bibnamefont
  {Singh}}, \bibinfo {author} {\bibfnamefont {R.}~\bibnamefont {Zeier}},
  \bibinfo {author} {\bibfnamefont {T.}~\bibnamefont {Calarco}},\ and\ \bibinfo
  {author} {\bibfnamefont {F.}~\bibnamefont {Motzoi}},\ }\bibfield  {title}
  {\bibinfo {title} {Compensating for nonlinear distortions in controlled
  quantum systems},\ }\href {https://doi.org/10.1103/PhysRevApplied.19.064067}
  {\bibfield  {journal} {\bibinfo  {journal} {Phys. Rev. Appl.}\ }\textbf
  {\bibinfo {volume} {19}},\ \bibinfo {pages} {064067} (\bibinfo {year}
  {2023})}\BibitemShut {NoStop}%
\bibitem [{\citenamefont {Jaeger}\ and\ \citenamefont
  {Hohenester}(2013)}]{jager2013}%
  \BibitemOpen
  \bibfield  {author} {\bibinfo {author} {\bibfnamefont {G.}~\bibnamefont
  {Jaeger}}\ and\ \bibinfo {author} {\bibfnamefont {U.}~\bibnamefont
  {Hohenester}},\ }\bibfield  {title} {\bibinfo {title} {Optimal quantum
  control of bose-einstein condensates in magnetic microtraps: Consideration of
  filter effects},\ }\href {https://doi.org/10.1103/PhysRevA.88.035601}
  {\bibfield  {journal} {\bibinfo  {journal} {Phys. Rev. A}\ }\textbf {\bibinfo
  {volume} {88}},\ \bibinfo {pages} {035601} (\bibinfo {year}
  {2013})}\BibitemShut {NoStop}%
\bibitem [{\citenamefont {Gustavsson}\ \emph {et~al.}(2013)\citenamefont
  {Gustavsson}, \citenamefont {Zwier}, \citenamefont {Bylander}, \citenamefont
  {Yan}, \citenamefont {Yoshihara}, \citenamefont {Nakamura}, \citenamefont
  {Orlando},\ and\ \citenamefont {Oliver}}]{Gustavsson2013}%
  \BibitemOpen
  \bibfield  {author} {\bibinfo {author} {\bibfnamefont {S.}~\bibnamefont
  {Gustavsson}}, \bibinfo {author} {\bibfnamefont {O.}~\bibnamefont {Zwier}},
  \bibinfo {author} {\bibfnamefont {J.}~\bibnamefont {Bylander}}, \bibinfo
  {author} {\bibfnamefont {F.}~\bibnamefont {Yan}}, \bibinfo {author}
  {\bibfnamefont {F.}~\bibnamefont {Yoshihara}}, \bibinfo {author}
  {\bibfnamefont {Y.}~\bibnamefont {Nakamura}}, \bibinfo {author}
  {\bibfnamefont {T.~P.}\ \bibnamefont {Orlando}},\ and\ \bibinfo {author}
  {\bibfnamefont {W.~D.}\ \bibnamefont {Oliver}},\ }\bibfield  {title}
  {\bibinfo {title} {Improving quantum gate fidelities by using a qubit to
  measure microwave pulse distortions},\ }\href
  {https://doi.org/10.1103/PhysRevLett.110.040502} {\bibfield  {journal}
  {\bibinfo  {journal} {Phys. Rev. Lett.}\ }\textbf {\bibinfo {volume} {110}},\
  \bibinfo {pages} {040502} (\bibinfo {year} {2013})}\BibitemShut {NoStop}%
\bibitem [{\citenamefont {Jerger}\ \emph {et~al.}(2019)\citenamefont {Jerger},
  \citenamefont {Kulikov}, \citenamefont {Vasselin},\ and\ \citenamefont
  {Fedorov}}]{Jerger2019}%
  \BibitemOpen
  \bibfield  {author} {\bibinfo {author} {\bibfnamefont {M.}~\bibnamefont
  {Jerger}}, \bibinfo {author} {\bibfnamefont {A.}~\bibnamefont {Kulikov}},
  \bibinfo {author} {\bibfnamefont {Z.}~\bibnamefont {Vasselin}},\ and\
  \bibinfo {author} {\bibfnamefont {A.}~\bibnamefont {Fedorov}},\ }\bibfield
  {title} {\bibinfo {title} {In situ characterization of qubit control lines: A
  qubit as a vector network analyzer},\ }\href
  {https://doi.org/10.1103/PhysRevLett.123.150501} {\bibfield  {journal}
  {\bibinfo  {journal} {Phys. Rev. Lett.}\ }\textbf {\bibinfo {volume} {123}},\
  \bibinfo {pages} {150501} (\bibinfo {year} {2019})}\BibitemShut {NoStop}%
\bibitem [{\citenamefont {Laz\ifmmode~\u{a}\else \u{a}\fi{}r}\ \emph
  {et~al.}(2023)\citenamefont {Laz\ifmmode~\u{a}\else \u{a}\fi{}r},
  \citenamefont {Ficheux}, \citenamefont {Herrmann}, \citenamefont {Remm},
  \citenamefont {Lacroix}, \citenamefont {Hellings}, \citenamefont {Swiadek},
  \citenamefont {Zanuz}, \citenamefont {Norris}, \citenamefont {Panah},
  \citenamefont {Flasby}, \citenamefont {Kerschbaum}, \citenamefont {Besse},
  \citenamefont {Eichler},\ and\ \citenamefont
  {Wallraff}}]{lazifmmode2023calibration}%
  \BibitemOpen
  \bibfield  {author} {\bibinfo {author} {\bibfnamefont {S.}~\bibnamefont
  {Laz\ifmmode~\u{a}\else \u{a}\fi{}r}}, \bibinfo {author} {\bibfnamefont
  {Q.}~\bibnamefont {Ficheux}}, \bibinfo {author} {\bibfnamefont
  {J.}~\bibnamefont {Herrmann}}, \bibinfo {author} {\bibfnamefont
  {A.}~\bibnamefont {Remm}}, \bibinfo {author} {\bibfnamefont {N.}~\bibnamefont
  {Lacroix}}, \bibinfo {author} {\bibfnamefont {C.}~\bibnamefont {Hellings}},
  \bibinfo {author} {\bibfnamefont {F.}~\bibnamefont {Swiadek}}, \bibinfo
  {author} {\bibfnamefont {D.~C.}\ \bibnamefont {Zanuz}}, \bibinfo {author}
  {\bibfnamefont {G.~J.}\ \bibnamefont {Norris}}, \bibinfo {author}
  {\bibfnamefont {M.~B.}\ \bibnamefont {Panah}}, \bibinfo {author}
  {\bibfnamefont {A.}~\bibnamefont {Flasby}}, \bibinfo {author} {\bibfnamefont
  {M.}~\bibnamefont {Kerschbaum}}, \bibinfo {author} {\bibfnamefont {J.-C.}\
  \bibnamefont {Besse}}, \bibinfo {author} {\bibfnamefont {C.}~\bibnamefont
  {Eichler}},\ and\ \bibinfo {author} {\bibfnamefont {A.}~\bibnamefont
  {Wallraff}},\ }\bibfield  {title} {\bibinfo {title} {Calibration of drive
  nonlinearity for arbitrary-angle single-qubit gates using error
  amplification},\ }\href {https://doi.org/10.1103/PhysRevApplied.20.024036}
  {\bibfield  {journal} {\bibinfo  {journal} {Phys. Rev. Appl.}\ }\textbf
  {\bibinfo {volume} {20}},\ \bibinfo {pages} {024036} (\bibinfo {year}
  {2023})}\BibitemShut {NoStop}%
\bibitem [{\citenamefont {Bruzewicz}\ \emph {et~al.}(2019)\citenamefont
  {Bruzewicz}, \citenamefont {Chiaverini}, \citenamefont {McConnell},\ and\
  \citenamefont {Sage}}]{bruzewicz2019trapped}%
  \BibitemOpen
  \bibfield  {author} {\bibinfo {author} {\bibfnamefont {C.~D.}\ \bibnamefont
  {Bruzewicz}}, \bibinfo {author} {\bibfnamefont {J.}~\bibnamefont
  {Chiaverini}}, \bibinfo {author} {\bibfnamefont {R.}~\bibnamefont
  {McConnell}},\ and\ \bibinfo {author} {\bibfnamefont {J.~M.}\ \bibnamefont
  {Sage}},\ }\bibfield  {title} {\bibinfo {title} {Trapped-ion quantum
  computing: Progress and challenges},\ }\href
  {https://doi.org/10.48550/arXiv.1904.04178} {\bibfield  {journal} {\bibinfo
  {journal} {Applied Physics Reviews}\ }\textbf {\bibinfo {volume} {6}}
  (\bibinfo {year} {2019})}\BibitemShut {NoStop}%
\bibitem [{\citenamefont {Harty}\ \emph {et~al.}(2016)\citenamefont {Harty},
  \citenamefont {Sepiol}, \citenamefont {Allcock}, \citenamefont {Ballance},
  \citenamefont {Tarlton},\ and\ \citenamefont {Lucas}}]{harty2016}%
  \BibitemOpen
  \bibfield  {author} {\bibinfo {author} {\bibfnamefont {T.~P.}\ \bibnamefont
  {Harty}}, \bibinfo {author} {\bibfnamefont {M.~A.}\ \bibnamefont {Sepiol}},
  \bibinfo {author} {\bibfnamefont {D.~T.~C.}\ \bibnamefont {Allcock}},
  \bibinfo {author} {\bibfnamefont {C.~J.}\ \bibnamefont {Ballance}}, \bibinfo
  {author} {\bibfnamefont {J.~E.}\ \bibnamefont {Tarlton}},\ and\ \bibinfo
  {author} {\bibfnamefont {D.~M.}\ \bibnamefont {Lucas}},\ }\bibfield  {title}
  {\bibinfo {title} {High-fidelity trapped-ion quantum logic using near-field
  microwaves},\ }\href {https://doi.org/10.1103/PhysRevLett.117.140501}
  {\bibfield  {journal} {\bibinfo  {journal} {Phys. Rev. Lett.}\ }\textbf
  {\bibinfo {volume} {117}},\ \bibinfo {pages} {140501} (\bibinfo {year}
  {2016})}\BibitemShut {NoStop}%
\bibitem [{\citenamefont {Bowler}\ \emph {et~al.}(2013)\citenamefont {Bowler},
  \citenamefont {Warring}, \citenamefont {Britton}, \citenamefont {Sawyer},\
  and\ \citenamefont {Amini}}]{bowler2013}%
  \BibitemOpen
  \bibfield  {author} {\bibinfo {author} {\bibfnamefont {R.}~\bibnamefont
  {Bowler}}, \bibinfo {author} {\bibfnamefont {U.}~\bibnamefont {Warring}},
  \bibinfo {author} {\bibfnamefont {J.~W.}\ \bibnamefont {Britton}}, \bibinfo
  {author} {\bibfnamefont {B.~C.}\ \bibnamefont {Sawyer}},\ and\ \bibinfo
  {author} {\bibfnamefont {J.}~\bibnamefont {Amini}},\ }\bibfield  {title}
  {\bibinfo {title} {{Arbitrary waveform generator for quantum information
  processing with trapped ions}},\ }\href {https://doi.org/10.1063/1.4795552}
  {\bibfield  {journal} {\bibinfo  {journal} {Review of Scientific
  Instruments}\ }\textbf {\bibinfo {volume} {84}},\ \bibinfo {pages} {033108}
  (\bibinfo {year} {2013})}\BibitemShut {NoStop}%
\bibitem [{\citenamefont {Poitzsch}\ \emph {et~al.}(1996)\citenamefont
  {Poitzsch}, \citenamefont {Bergquist}, \citenamefont {Itano},\ and\
  \citenamefont {Wineland}}]{poitzsch1996cryogenic}%
  \BibitemOpen
  \bibfield  {author} {\bibinfo {author} {\bibfnamefont {M.~E.}\ \bibnamefont
  {Poitzsch}}, \bibinfo {author} {\bibfnamefont {J.~C.}\ \bibnamefont
  {Bergquist}}, \bibinfo {author} {\bibfnamefont {W.~M.}\ \bibnamefont
  {Itano}},\ and\ \bibinfo {author} {\bibfnamefont {D.~J.}\ \bibnamefont
  {Wineland}},\ }\bibfield  {title} {\bibinfo {title} {Cryogenic linear ion
  trap for accurate spectroscopy},\ }\href
  {https://tsapps.nist.gov/publication/get_pdf.cfm?pub_id=105056} {\bibfield
  {journal} {\bibinfo  {journal} {Review of Scientific Instruments}\ }\textbf
  {\bibinfo {volume} {67}},\ \bibinfo {pages} {129} (\bibinfo {year}
  {1996})}\BibitemShut {NoStop}%
\bibitem [{\citenamefont {Kristen}\ \emph {et~al.}(2020)\citenamefont
  {Kristen}, \citenamefont {Schneider}, \citenamefont {Stehli}, \citenamefont
  {Wolz}, \citenamefont {Danilin}, \citenamefont {Ku}, \citenamefont {Long},
  \citenamefont {Wu}, \citenamefont {Lake}, \citenamefont {Pappas} \emph
  {et~al.}}]{kristen2020amplitude}%
  \BibitemOpen
  \bibfield  {author} {\bibinfo {author} {\bibfnamefont {M.}~\bibnamefont
  {Kristen}}, \bibinfo {author} {\bibfnamefont {A.}~\bibnamefont {Schneider}},
  \bibinfo {author} {\bibfnamefont {A.}~\bibnamefont {Stehli}}, \bibinfo
  {author} {\bibfnamefont {T.}~\bibnamefont {Wolz}}, \bibinfo {author}
  {\bibfnamefont {S.}~\bibnamefont {Danilin}}, \bibinfo {author} {\bibfnamefont
  {H.~S.}\ \bibnamefont {Ku}}, \bibinfo {author} {\bibfnamefont
  {J.}~\bibnamefont {Long}}, \bibinfo {author} {\bibfnamefont {X.}~\bibnamefont
  {Wu}}, \bibinfo {author} {\bibfnamefont {R.}~\bibnamefont {Lake}}, \bibinfo
  {author} {\bibfnamefont {D.~P.}\ \bibnamefont {Pappas}}, \emph {et~al.},\
  }\bibfield  {title} {\bibinfo {title} {Amplitude and frequency sensing of
  microwave fields with a superconducting transmon qudit},\ }\href
  {https://doi.org/10.1038/s41534-020-00287-w} {\bibfield  {journal} {\bibinfo
  {journal} {npj Quantum Information}\ }\textbf {\bibinfo {volume} {6}},\
  \bibinfo {pages} {57} (\bibinfo {year} {2020})}\BibitemShut {NoStop}%
\bibitem [{\citenamefont {Ockeloen}\ \emph {et~al.}(2013)\citenamefont
  {Ockeloen}, \citenamefont {Schmied}, \citenamefont {Riedel},\ and\
  \citenamefont {Treutlein}}]{Ockeloen2013}%
  \BibitemOpen
  \bibfield  {author} {\bibinfo {author} {\bibfnamefont {C.~F.}\ \bibnamefont
  {Ockeloen}}, \bibinfo {author} {\bibfnamefont {R.}~\bibnamefont {Schmied}},
  \bibinfo {author} {\bibfnamefont {M.~F.}\ \bibnamefont {Riedel}},\ and\
  \bibinfo {author} {\bibfnamefont {P.}~\bibnamefont {Treutlein}},\ }\bibfield
  {title} {\bibinfo {title} {Quantum metrology with a scanning probe atom
  interferometer},\ }\href {https://doi.org/10.1103/PhysRevLett.111.143001}
  {\bibfield  {journal} {\bibinfo  {journal} {Phys. Rev. Lett.}\ }\textbf
  {\bibinfo {volume} {111}},\ \bibinfo {pages} {143001} (\bibinfo {year}
  {2013})}\BibitemShut {NoStop}%
\bibitem [{\citenamefont {Bohi}\ \emph {et~al.}(2010)\citenamefont {Bohi},
  \citenamefont {Riedel}, \citenamefont {Hansch},\ and\ \citenamefont
  {Treutlein}}]{Bohi2010}%
  \BibitemOpen
  \bibfield  {author} {\bibinfo {author} {\bibfnamefont {P.}~\bibnamefont
  {Bohi}}, \bibinfo {author} {\bibfnamefont {M.~F.}\ \bibnamefont {Riedel}},
  \bibinfo {author} {\bibfnamefont {T.~W.}\ \bibnamefont {Hansch}},\ and\
  \bibinfo {author} {\bibfnamefont {P.}~\bibnamefont {Treutlein}},\ }\bibfield
  {title} {\bibinfo {title} {{Imaging of microwave fields using ultracold
  atoms}},\ }\href {https://doi.org/10.1063/1.3470591} {\bibfield  {journal}
  {\bibinfo  {journal} {Applied Physics Letters}\ }\textbf {\bibinfo {volume}
  {97}},\ \bibinfo {pages} {051101} (\bibinfo {year} {2010})},\ \Eprint
  {https://arxiv.org/abs/https://pubs.aip.org/aip/apl/article-pdf/doi/10.1063/1.3470591/13581421/051101\_1\_online.pdf}
  {https://pubs.aip.org/aip/apl/article-pdf/doi/10.1063/1.3470591/13581421/051101\_1\_online.pdf}
  \BibitemShut {NoStop}%
\bibitem [{\citenamefont {Ospelkaus}\ \emph {et~al.}(2008)\citenamefont
  {Ospelkaus}, \citenamefont {Langer}, \citenamefont {Amini}, \citenamefont
  {Brown}, \citenamefont {Leibfried},\ and\ \citenamefont
  {Wineland}}]{ospelkaus2008}%
  \BibitemOpen
  \bibfield  {author} {\bibinfo {author} {\bibfnamefont {C.}~\bibnamefont
  {Ospelkaus}}, \bibinfo {author} {\bibfnamefont {C.~E.}\ \bibnamefont
  {Langer}}, \bibinfo {author} {\bibfnamefont {J.~M.}\ \bibnamefont {Amini}},
  \bibinfo {author} {\bibfnamefont {K.~R.}\ \bibnamefont {Brown}}, \bibinfo
  {author} {\bibfnamefont {D.}~\bibnamefont {Leibfried}},\ and\ \bibinfo
  {author} {\bibfnamefont {D.~J.}\ \bibnamefont {Wineland}},\ }\bibfield
  {title} {\bibinfo {title} {Trapped-ion quantum logic gates based on
  oscillating magnetic fields},\ }\href
  {https://doi.org/10.1103/PhysRevLett.101.090502} {\bibfield  {journal}
  {\bibinfo  {journal} {Phys. Rev. Lett.}\ }\textbf {\bibinfo {volume} {101}},\
  \bibinfo {pages} {090502} (\bibinfo {year} {2008})}\BibitemShut {NoStop}%
\bibitem [{\citenamefont {Warring}\ \emph {et~al.}(2013)\citenamefont
  {Warring}, \citenamefont {Ospelkaus}, \citenamefont {Colombe}, \citenamefont
  {Brown}, \citenamefont {Amini}, \citenamefont {Carsjens}, \citenamefont
  {Leibfried},\ and\ \citenamefont {Wineland}}]{Warring2013Phase}%
  \BibitemOpen
  \bibfield  {author} {\bibinfo {author} {\bibfnamefont {U.}~\bibnamefont
  {Warring}}, \bibinfo {author} {\bibfnamefont {C.}~\bibnamefont {Ospelkaus}},
  \bibinfo {author} {\bibfnamefont {Y.}~\bibnamefont {Colombe}}, \bibinfo
  {author} {\bibfnamefont {K.~R.}\ \bibnamefont {Brown}}, \bibinfo {author}
  {\bibfnamefont {J.~M.}\ \bibnamefont {Amini}}, \bibinfo {author}
  {\bibfnamefont {M.}~\bibnamefont {Carsjens}}, \bibinfo {author}
  {\bibfnamefont {D.}~\bibnamefont {Leibfried}},\ and\ \bibinfo {author}
  {\bibfnamefont {D.~J.}\ \bibnamefont {Wineland}},\ }\bibfield  {title}
  {\bibinfo {title} {Techniques for microwave near-field quantum control of
  trapped ions},\ }\href {https://doi.org/10.1103/PhysRevA.87.013437}
  {\bibfield  {journal} {\bibinfo  {journal} {Phys. Rev. A}\ }\textbf {\bibinfo
  {volume} {87}},\ \bibinfo {pages} {013437} (\bibinfo {year}
  {2013})}\BibitemShut {NoStop}%
\bibitem [{\citenamefont {Harty}\ \emph {et~al.}(2014)\citenamefont {Harty},
  \citenamefont {Allcock}, \citenamefont {Ballance}, \citenamefont {Guidoni},
  \citenamefont {Janacek}, \citenamefont {Linke}, \citenamefont {Stacey},\ and\
  \citenamefont {Lucas}}]{harty2014}%
  \BibitemOpen
  \bibfield  {author} {\bibinfo {author} {\bibfnamefont {T.~P.}\ \bibnamefont
  {Harty}}, \bibinfo {author} {\bibfnamefont {D.~T.~C.}\ \bibnamefont
  {Allcock}}, \bibinfo {author} {\bibfnamefont {C.~J.}\ \bibnamefont
  {Ballance}}, \bibinfo {author} {\bibfnamefont {L.}~\bibnamefont {Guidoni}},
  \bibinfo {author} {\bibfnamefont {H.~A.}\ \bibnamefont {Janacek}}, \bibinfo
  {author} {\bibfnamefont {N.~M.}\ \bibnamefont {Linke}}, \bibinfo {author}
  {\bibfnamefont {D.~N.}\ \bibnamefont {Stacey}},\ and\ \bibinfo {author}
  {\bibfnamefont {D.~M.}\ \bibnamefont {Lucas}},\ }\bibfield  {title} {\bibinfo
  {title} {High-fidelity preparation, gates, memory, and readout of a
  trapped-ion quantum bit},\ }\href
  {https://doi.org/10.1103/PhysRevLett.113.220501} {\bibfield  {journal}
  {\bibinfo  {journal} {Phys. Rev. Lett.}\ }\textbf {\bibinfo {volume} {113}},\
  \bibinfo {pages} {220501} (\bibinfo {year} {2014})}\BibitemShut {NoStop}%
\bibitem [{\citenamefont {Zarantonello}\ \emph {et~al.}(2019)\citenamefont
  {Zarantonello}, \citenamefont {Hahn}, \citenamefont {Morgner}, \citenamefont
  {Schulte}, \citenamefont {Bautista-Salvador}, \citenamefont {Werner},
  \citenamefont {Hammerer},\ and\ \citenamefont
  {Ospelkaus}}]{zarantonello2019}%
  \BibitemOpen
  \bibfield  {author} {\bibinfo {author} {\bibfnamefont {G.}~\bibnamefont
  {Zarantonello}}, \bibinfo {author} {\bibfnamefont {H.}~\bibnamefont {Hahn}},
  \bibinfo {author} {\bibfnamefont {J.}~\bibnamefont {Morgner}}, \bibinfo
  {author} {\bibfnamefont {M.}~\bibnamefont {Schulte}}, \bibinfo {author}
  {\bibfnamefont {A.}~\bibnamefont {Bautista-Salvador}}, \bibinfo {author}
  {\bibfnamefont {R.~F.}\ \bibnamefont {Werner}}, \bibinfo {author}
  {\bibfnamefont {K.}~\bibnamefont {Hammerer}},\ and\ \bibinfo {author}
  {\bibfnamefont {C.}~\bibnamefont {Ospelkaus}},\ }\bibfield  {title} {\bibinfo
  {title} {Robust and resource-efficient microwave near-field entangling
  $^{9}{\mathrm{be}}^{+}$ gate},\ }\href
  {https://doi.org/10.1103/PhysRevLett.123.260503} {\bibfield  {journal}
  {\bibinfo  {journal} {Phys. Rev. Lett.}\ }\textbf {\bibinfo {volume} {123}},\
  \bibinfo {pages} {260503} (\bibinfo {year} {2019})}\BibitemShut {NoStop}%
\bibitem [{\citenamefont {Srinivas}\ \emph {et~al.}(2021)\citenamefont
  {Srinivas}, \citenamefont {Burd}, \citenamefont {Knaack}, \citenamefont
  {Sutherland}, \citenamefont {Kwiatkowski}, \citenamefont {Glancy},
  \citenamefont {Knill}, \citenamefont {Wineland}, \citenamefont {Leibfried},
  \citenamefont {Wilson}, \citenamefont {Allcock},\ and\ \citenamefont
  {Slichter}}]{srinivas2021}%
  \BibitemOpen
  \bibfield  {author} {\bibinfo {author} {\bibfnamefont {R.}~\bibnamefont
  {Srinivas}}, \bibinfo {author} {\bibfnamefont {S.~C.}\ \bibnamefont {Burd}},
  \bibinfo {author} {\bibfnamefont {H.~M.}\ \bibnamefont {Knaack}}, \bibinfo
  {author} {\bibfnamefont {R.~T.}\ \bibnamefont {Sutherland}}, \bibinfo
  {author} {\bibfnamefont {A.}~\bibnamefont {Kwiatkowski}}, \bibinfo {author}
  {\bibfnamefont {S.}~\bibnamefont {Glancy}}, \bibinfo {author} {\bibfnamefont
  {E.}~\bibnamefont {Knill}}, \bibinfo {author} {\bibfnamefont {D.~J.}\
  \bibnamefont {Wineland}}, \bibinfo {author} {\bibfnamefont {D.}~\bibnamefont
  {Leibfried}}, \bibinfo {author} {\bibfnamefont {A.~C.}\ \bibnamefont
  {Wilson}}, \bibinfo {author} {\bibfnamefont {D.~T.~C.}\ \bibnamefont
  {Allcock}},\ and\ \bibinfo {author} {\bibfnamefont {D.~H.}\ \bibnamefont
  {Slichter}},\ }\bibfield  {title} {\bibinfo {title} {High-fidelity laser-free
  universal control of trapped ion qubits},\ }\href
  {https://doi.org/10.1038/s41586-021-03809-4} {\bibfield  {journal} {\bibinfo
  {journal} {Nature}\ }\textbf {\bibinfo {volume} {597}},\ \bibinfo {pages}
  {209–213} (\bibinfo {year} {2021})}\BibitemShut {NoStop}%
\bibitem [{\citenamefont {Leu}\ \emph {et~al.}(2023)\citenamefont {Leu},
  \citenamefont {Gely}, \citenamefont {Weber}, \citenamefont {Smith},
  \citenamefont {Nadlinger},\ and\ \citenamefont {Lucas}}]{leu2023}%
  \BibitemOpen
  \bibfield  {author} {\bibinfo {author} {\bibfnamefont {A.~D.}\ \bibnamefont
  {Leu}}, \bibinfo {author} {\bibfnamefont {M.~F.}\ \bibnamefont {Gely}},
  \bibinfo {author} {\bibfnamefont {M.~A.}\ \bibnamefont {Weber}}, \bibinfo
  {author} {\bibfnamefont {M.~C.}\ \bibnamefont {Smith}}, \bibinfo {author}
  {\bibfnamefont {D.~P.}\ \bibnamefont {Nadlinger}},\ and\ \bibinfo {author}
  {\bibfnamefont {D.~M.}\ \bibnamefont {Lucas}},\ }\bibfield  {title} {\bibinfo
  {title} {Fast, high-fidelity addressed single-qubit gates using efficient
  composite pulse sequences},\ }\href
  {https://doi.org/10.1103/PhysRevLett.131.120601} {\bibfield  {journal}
  {\bibinfo  {journal} {Phys. Rev. Lett.}\ }\textbf {\bibinfo {volume} {131}},\
  \bibinfo {pages} {120601} (\bibinfo {year} {2023})}\BibitemShut {NoStop}%
\bibitem [{\citenamefont {Weber}\ \emph {et~al.}(2022)\citenamefont {Weber},
  \citenamefont {Löschnauer}, \citenamefont {Wolf}, \citenamefont {Gely},
  \citenamefont {Hanley}, \citenamefont {Goodwin}, \citenamefont {Ballance},
  \citenamefont {Harty},\ and\ \citenamefont {Lucas}}]{weber2022cryogenic}%
  \BibitemOpen
  \bibfield  {author} {\bibinfo {author} {\bibfnamefont {M.~A.}\ \bibnamefont
  {Weber}}, \bibinfo {author} {\bibfnamefont {C.}~\bibnamefont {Löschnauer}},
  \bibinfo {author} {\bibfnamefont {J.}~\bibnamefont {Wolf}}, \bibinfo {author}
  {\bibfnamefont {M.~F.}\ \bibnamefont {Gely}}, \bibinfo {author}
  {\bibfnamefont {R.~K.}\ \bibnamefont {Hanley}}, \bibinfo {author}
  {\bibfnamefont {J.~F.}\ \bibnamefont {Goodwin}}, \bibinfo {author}
  {\bibfnamefont {C.~J.}\ \bibnamefont {Ballance}}, \bibinfo {author}
  {\bibfnamefont {T.~P.}\ \bibnamefont {Harty}},\ and\ \bibinfo {author}
  {\bibfnamefont {D.~M.}\ \bibnamefont {Lucas}},\ }\href@noop {} {\bibinfo
  {title} {Cryogenic ion trap system for high-fidelity near-field
  microwave-driven quantum logic}} (\bibinfo {year} {2022}),\ \Eprint
  {https://arxiv.org/abs/2207.11364} {arXiv:2207.11364 [quant-ph]} \BibitemShut
  {NoStop}%
\bibitem [{\citenamefont {Knill}\ \emph {et~al.}(2008)\citenamefont {Knill},
  \citenamefont {Leibfried}, \citenamefont {Reichle}, \citenamefont {Britton},
  \citenamefont {Blakestad}, \citenamefont {Jost}, \citenamefont {Langer},
  \citenamefont {Ozeri}, \citenamefont {Seidelin},\ and\ \citenamefont
  {Wineland}}]{Knill2008RBM}%
  \BibitemOpen
  \bibfield  {author} {\bibinfo {author} {\bibfnamefont {E.}~\bibnamefont
  {Knill}}, \bibinfo {author} {\bibfnamefont {D.}~\bibnamefont {Leibfried}},
  \bibinfo {author} {\bibfnamefont {R.}~\bibnamefont {Reichle}}, \bibinfo
  {author} {\bibfnamefont {J.}~\bibnamefont {Britton}}, \bibinfo {author}
  {\bibfnamefont {R.~B.}\ \bibnamefont {Blakestad}}, \bibinfo {author}
  {\bibfnamefont {J.~D.}\ \bibnamefont {Jost}}, \bibinfo {author}
  {\bibfnamefont {C.}~\bibnamefont {Langer}}, \bibinfo {author} {\bibfnamefont
  {R.}~\bibnamefont {Ozeri}}, \bibinfo {author} {\bibfnamefont
  {S.}~\bibnamefont {Seidelin}},\ and\ \bibinfo {author} {\bibfnamefont
  {D.~J.}\ \bibnamefont {Wineland}},\ }\bibfield  {title} {\bibinfo {title}
  {Randomized benchmarking of quantum gates},\ }\href
  {https://doi.org/10.1103/PhysRevA.77.012307} {\bibfield  {journal} {\bibinfo
  {journal} {Phys. Rev. A}\ }\textbf {\bibinfo {volume} {77}},\ \bibinfo
  {pages} {012307} (\bibinfo {year} {2008})}\BibitemShut {NoStop}%
\bibitem [{\citenamefont {Takeda}\ \emph {et~al.}(2016)\citenamefont {Takeda},
  \citenamefont {Kamioka}, \citenamefont {Otsuka}, \citenamefont {Yoneda},
  \citenamefont {Nakajima}, \citenamefont {Delbecq}, \citenamefont {Amaha},
  \citenamefont {Allison}, \citenamefont {Kodera}, \citenamefont {Oda},\ and\
  \citenamefont {Tarucha}}]{Takeda2016}%
  \BibitemOpen
  \bibfield  {author} {\bibinfo {author} {\bibfnamefont {K.}~\bibnamefont
  {Takeda}}, \bibinfo {author} {\bibfnamefont {J.}~\bibnamefont {Kamioka}},
  \bibinfo {author} {\bibfnamefont {T.}~\bibnamefont {Otsuka}}, \bibinfo
  {author} {\bibfnamefont {J.}~\bibnamefont {Yoneda}}, \bibinfo {author}
  {\bibfnamefont {T.}~\bibnamefont {Nakajima}}, \bibinfo {author}
  {\bibfnamefont {M.~R.}\ \bibnamefont {Delbecq}}, \bibinfo {author}
  {\bibfnamefont {S.}~\bibnamefont {Amaha}}, \bibinfo {author} {\bibfnamefont
  {G.}~\bibnamefont {Allison}}, \bibinfo {author} {\bibfnamefont
  {T.}~\bibnamefont {Kodera}}, \bibinfo {author} {\bibfnamefont
  {S.}~\bibnamefont {Oda}},\ and\ \bibinfo {author} {\bibfnamefont
  {S.}~\bibnamefont {Tarucha}},\ }\bibfield  {title} {\bibinfo {title} {A
  fault-tolerant addressable spin qubit in a natural silicon quantum dot},\
  }\href {https://doi.org/10.1126/sciadv.1600694} {\bibfield  {journal}
  {\bibinfo  {journal} {Science Advances}\ }\textbf {\bibinfo {volume} {2}},\
  \bibinfo {pages} {e1600694} (\bibinfo {year} {2016})},\ \Eprint
  {https://arxiv.org/abs/https://www.science.org/doi/pdf/10.1126/sciadv.1600694}
  {https://www.science.org/doi/pdf/10.1126/sciadv.1600694} \BibitemShut
  {NoStop}%
\end{thebibliography}%

\FloatBarrier
\clearpage
\onecolumngrid
\begin{center}
{\Large \textbf{Supplementary information}}
\end{center}
\makeatletter
   \renewcommand\l@section{\@dottedtocline{2}{1.5em}{2em}}
   \renewcommand\l@subsection{\@dottedtocline{2}{3.5em}{2em}}
   \renewcommand\l@subsubsection{\@dottedtocline{2}{5.5em}{2em}}
\makeatother
\let\addcontentsline\oldaddcontentsline

\renewcommand{\thesection}{\arabic{section}}

\twocolumngrid

\let\oldaddcontentsline\addcontentsline
\renewcommand{\addcontentsline}[3]{}
\let\addcontentsline\oldaddcontentsline
\renewcommand{\theequation}{S\arabic{equation}}
\renewcommand{\thefigure}{S\arabic{figure}}
\renewcommand{\thetable}{S\arabic{table}}
\renewcommand{\thesection}{S\arabic{section}}
\setcounter{figure}{0}
\setcounter{equation}{0}
\setcounter{section}{0}

\newcolumntype{C}[1]{>{\centering\arraybackslash}p{#1}}
\newcolumntype{L}[1]{>{\raggedright\arraybackslash}p{#1}}



\section{Theory}
\label{sec:theory}

\subsection{ZYXYZ decomposition}
\label{sec:ZYXYZ_decomposition}

The Hamiltonian describing MW driving of the qubit with a time-varying phase writes
\begin{equation}
    \hat H (t) = \hbar\underbrace{ \Omega(t) \cos(\phi(t))}_{\Omega_x(t)}\hat \sigma_x/2+ \hbar \underbrace{\Omega(t)\sin(\phi(t))}_{\Omega_y(t)}\hat\sigma_y/2\ ,
    \label{eq:ZYXYZ_H}
\end{equation}
in the interaction picture, where terms rotating at the qubit frequency have been neglected.
Here $\hbar$ is the reduced Planck constant, $\hat\sigma_{x,y}$ are Pauli operators acting on the qubit states, and $\Omega(t),\phi(t)$ are the Rabi-frequency and phase of the drive.
The variations in phase are assumed to arise from changes in amplitude, such that, without loss of generality, we can set $\phi(t)=0$ during the ``plateau'' of the pulse.
The small phase variations then perturb the pulse -- during the ramping -- away from the ideal $x$-rotation.
This Hamiltonian drives the following unitary evolution during ramping
\begin{equation}
    \hat U_\text{ramp}=\exp[-i(r_x\hat\sigma_x+r_y\hat\sigma_y)/2]\ , 
\end{equation}
where $r_{x,y}=\int_0^{t_\text{ramp}}\Omega_{x,y}(t)dt$.
Upon ramp up, we decompose the unitary into $z$-, $y$- then $x$- rotation described by
\begin{equation}
    \hat U_\text{ramp}\simeq\exp[-i\alpha_x\hat\sigma_x/2]\exp[-i\alpha_y\hat\sigma_y/2]\exp[-i\alpha_z\hat\sigma_z/2]\ ,
\end{equation}
where $\alpha_{x,y,z}$ are Tait-Bryan angles.
Whilst the decomposition can be exact, we can assume small phase deviations $|\phi(t)|\ll 1$, implying $|r_y|\ll 1$, to simplify the expression for the rotation angles, given to first order by
\begin{equation}
    \begin{aligned}
        \alpha_x &= r_x\\
        \alpha_y &= r_y \frac{\sin(r_x)}{r_x} \\
        \alpha_z &= -r_y \frac{1-\cos(r_x)}{r_x}\ .
    \end{aligned}
\end{equation}
Upon ramp-down, assuming a symmetric pulse $\int_{T-t_\text{ramp}}^T \Omega_{x,y}(t)dt=r_{x,y}$, we can decompose the unitary into a $x$-, $y$- then $z$- rotation described by
\begin{equation}
    \hat U_\text{ramp}\simeq\exp[-i(-\alpha_z)\hat\sigma_z/2]\exp[-i\alpha_y\hat\sigma_y/2]\exp[-i\alpha_x\hat\sigma_x/2]\ .
\end{equation}
The $z$-rotation angle has flipped sign here because of the change in the order of rotations with respect to the ramp-up.
When constructing a train of pulses, the inter-pulse z-rotations of subsequent pulses cancel due to this sign flip.
The remaining z-rotations at the beginning and end of the pulse train is negligeable, in addition to being irrelevant when preparing and measuring the qubit state in the computational basis.
The $x$-rotations will dominate in the ramping portion of the pulse, and can be combined with the unitary describing the ``plateau'' portion of the pulse, resulting in the effective description of the pulse shown in Fig~\ref{fig:fig2}(a).
Note that, given our definition of $\phi(t)$, and our assumption of positive proportionality factor between phase and amplitude in Fig.~\ref{fig:fig2}, $r_y<0$ in the trajectories of Fig.~\ref{fig:fig2}(b,c).

\subsection{Perturbation theory}
\label{sec:perturbation_theory}

\subsubsection{Problem statement}
To describe the data shown in Fig.~\ref{fig:fig3}(b), we introduce the relation between amplitude and phase $\phi(t) = \phi' (a(t)-A)$.
The Hamiltonian introduced in Eq.~(\ref{eq:ZYXYZ_H}) then writes
\begin{equation}
    \begin{aligned}
    \hat H (t) =& \hbar\Omega a(t)\big[ \cos\left(\phi' \left(a(t)-A\right)\right) \hat \sigma_x/2\\
     &+ \sin\left(\phi' \left(a(t)-A\right)\right) \hat \sigma_y/2 \big]\ .
\end{aligned}
\end{equation}
Here $a(t)$ describes the temporal profile and the amplitude of the pulse: we assume $\sin^2$ ramping over a duration $t_\text{ramp}$ at both the beginning and end of the pulse, and a plateau of amplitude $a(t)=A$.
The quantity $\Omega=2\pi/(T-t_\text{ramp})$ can be seen as a normalization factor which ensures that 
\begin{equation}
    \int_0^{T} \Omega a(t)dt = 2\pi A\ .
\end{equation} 
We consider a small deviation $\epsilon$ from a pulse $a_{2\pi}$ generating $2\pi$ rotations in the absence of phase variations ($\int_0^{T}\Omega a_{2\pi}(t)dt=2\pi$), such that the pulse profile writes
\begin{equation}
    a(t) = a_{2\pi}(t)(1+\epsilon)\ .
\end{equation}
In the limit $|\epsilon|,|\phi'| \ll 1$, the parameter regime explored in Fig.~\ref{fig:fig3}(b), the Hamiltonian can be approximated as
\begin{equation}
\begin{aligned}
    \hat H (t) &\simeq H_0(t)+H_1(t)\ ,\\
    \hat H_0(t) & = \hbar\Omega a_{2\pi}(t) \hat \sigma_x/2\ ,\\
    \hat H_1(t) & = \hbar\Omega \left(\epsilon a_{2\pi}(t) \hat \sigma_x/2 + \phi' a_{2\pi}(t)(a_{2\pi}(t)-1) \hat \sigma_y/2 \right)\ .
\end{aligned}
\end{equation}

\subsubsection{Perturbative calculations}
We aim to write the unitary evolution using first order perturbation theory:
\begin{equation}
    \begin{aligned}
    \hat U(0,T) &\simeq \hat U_0(0,T)+\hat U_1(0,T)\ ,\\
    \hat U_0(t_0,t) &= \exp \left[-\frac {i}{\hbar}\int_{t_0}^{t}\hat H_0 (\tau)d\tau\right]\\
    \hat U_1(0,T) &= -\frac{i}{\hbar}\hat U_0(0,T)\int_0^T \hat U_0(t,0) \hat H_1(t) \hat U_0(0,t) dt\ .
    \label{eq:S1B_perturbative_expansion}
\end{aligned}
\end{equation}
Since textbooks tend to focus on the case where $H_0$ is time-independent, we have not been able to find a proof in the literature of this exact perturbative expansion of the unitary evolution operator.
We therefore include a proof of (\ref{eq:S1B_perturbative_expansion}) in Sec.~\ref{sec:perturbative_expansion}.
The 0-th order term drives a rotation of angle $\theta(t)$ about the $x$- axis of the Bloch sphere:
\begin{equation}
    \begin{aligned}
        \hat U_0(0,t) &= \begin{pmatrix}
            \cos\left[\theta(t)/2\right] & -i\sin\left[\theta(t)/2\right]\\
            -i\sin\left[\theta(t)/2\right] & \cos\left[\theta(t)/2\right]\\
        \end{pmatrix},\\
        \hat U_0(t,0) &= \begin{pmatrix}
            \cos\left[\theta(t)/2\right] & i\sin\left[\theta(t)/2\right]\\
            i\sin\left[\theta(t)/2\right] & \cos\left[\theta(t)/2\right]\\
        \end{pmatrix},\\
            \theta(t) &= \int_0^{t} \Omega a_{2\pi}(\tau)d\tau\ .
    \end{aligned}
\end{equation}
Here the definition of $a_{2\pi}$ means that $\theta(T)=2\pi$ and $\hat U_0(0,T)=- \mathds{1}$.
Expressing $\hat U_0$ in this form is helpful when computing the first order term
\begin{equation}
    \begin{aligned}
    \hat U_1(0,T) &= i \epsilon A_x \hat \sigma_x/2+ i \phi' A_y \hat \sigma_y/2\ ,\\
     A_x &= \int_0^T \Omega a_{2\pi}(t) dt =2\pi\ ,\\
    A_y &= \int_0^T \Omega a_{2\pi}(t)(a_{2\pi}(t)-1)\cos(\theta(t)) dt\ .
    \end{aligned}
\end{equation}

To derive this expression, we have relied on the fact that $\int_0^T\sin(\theta(t))a(t)(a_{2\pi}(t)-1)=0$ since $\sin(\theta(t))$ and $a(t)$ are antisymmetric and symmetric respectively about $t=T/2$.
The total unitary in first order perturbation theory is then (up to a global phase)
\begin{equation}
    \begin{aligned}
        \hat U(0,T) &\simeq \mathds{1}-i \epsilon A_x \hat \sigma_x/2- i \phi' A_y \hat \sigma_y/2\\
        &\simeq \exp\left[-i\epsilon A_x \hat \sigma_x/2 -i \phi' A_y \hat \sigma_y/2\right]
    \label{eq:S_unitary_single_pulse}
    \end{aligned}
\end{equation}
where the exponential form is equivalent to first order in $\epsilon,\phi'$.
The exponential form makes the cumulative effect of multiple pulses easier to calculate:
\begin{equation}
        \hat U(0,T)^N \simeq \exp\left[-iN\left(\epsilon A_x \hat \sigma_x/2 +\phi' A_y \hat \sigma_y/2\right)\right]\ .
\end{equation}
This unitary corresponds to Rabi-oscillations where the probability of recovering the initial qubit state $\ket{0}$ is
\begin{equation}
    P_{\ket{0}} = \frac12\left(1+\cos\left(N\sqrt{(\epsilon A_x)^2+(\phi' A_y)^2}\right)\right)\ .
    \label{eq:S_P0}
\end{equation}

\subsubsection{Discussion}
This expression can be further Taylor expanded to explain why the loss of contrast only occurs in our data at $A=1$.
At $\epsilon=0$ ($A=1$), the probability $P_{\ket{0}}$ is approximately
\begin{equation}
    P_{\ket{0}} \simeq 1-\frac14\left(N\phi' A_y\right)^2\ ,
    \label{eq:S_sensitivity_1}
\end{equation}
showing a \textit{quadratic} dependence on phase variations.
At a pulse amplitude $\epsilon = 1/N$ -- corresponding to the next $P_{\ket{0}}\simeq1$ point at $A = (N+1)/N$ in Figs.~\ref{fig:fig1},\ref{fig:fig3} -- the condition $\epsilon A_x=2\pi/N\gg|\phi' A_y|$ is satisfied, such that Eq.(\ref{eq:S_P0}) can be written as
\begin{equation}
    \begin{aligned}
        P_{\ket{0}} &\simeq \frac12\left(1+\cos\left(N\epsilon A_x\left[1+\frac12\frac{(\phi' A_y)^2}{(\epsilon A_x)^2}\right]\right)\right)\\
        &\simeq \frac12\left(1+\cos\left(2\pi+\frac{(N\phi' A_y)^2}{4\pi}\right)\right)\\
        &\simeq 1-\frac1{64\pi^2}\left(N\phi' A_y\right)^4\ ,
        \label{eq:S_sensitivity_quartic}
    \end{aligned}
\end{equation}
showing a much smaller \textit{quartic} dependence on phase variations, explaining why loss of contrast only occurs around $A=1$ in our measurements.

\subsubsection{Comparisom to data}
\begin{figure}[h!]
    \centering
    \includegraphics[width=0.45\textwidth]{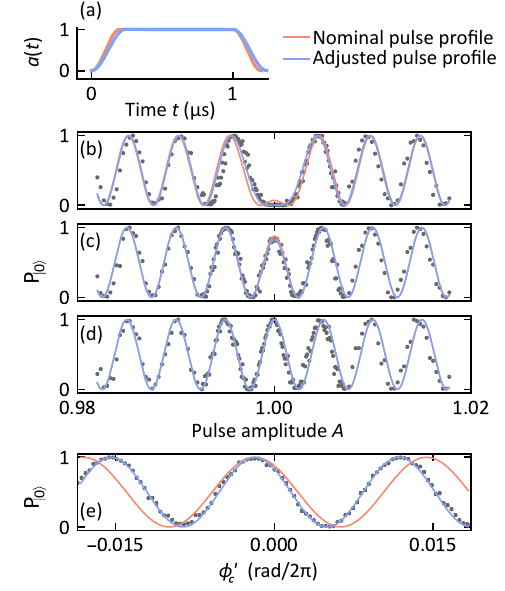}
    \caption{
        \textbf{Comparing perturbation theory and experimental data}
        (a) The nominal pulse profile and a slightly adjusted one are used here in computing theoretical predictions from perturbation theory (Eq.~\ref{eq:S_P0})
        (b-e) Line-cuts featured in Fig.~\ref{fig:fig3} are shown again here (dots), and compared to perturbation theory (lines).
        An adjustment of the pulse profile thus results in a better agreement between data and perturbation theory, which is not surprising as the nominal pulse profile (programmed into our AWG) is likely to be distorted as it propagates through the microwave delivery chain.
    }
    \label{fig:perturbation_vs_data}
\end{figure}

Plotting $P_{\ket{0}}$ would reproduce the concentric circles of Fig.~\ref{fig:fig3}(b) where $\epsilon$ and $\phi'$ are varied in the $x$ and $y$ axes respectively.
To quantitatively assess the results of perturbation theory, pulse amplitude sweeps are compared to Eq.~(\ref{eq:S_P0}) in Fig.~\ref{fig:perturbation_vs_data}, first with the nominal temporal profile generated by the AWG. 
The temporal pulse profile has not been independently characterized, and is likely to vary from it's nominal shape when generated and as it propagates through the microwave delivery chain.
By slightly varying the pulse ramp time in the perturbative model, better agreement to data is reached.

\subsection{Perturbative expansion of $\hat U$}
\label{sec:perturbative_expansion}

Here we provide proof of the perturbative expansion introduced in Eq.~(\ref{eq:S1B_perturbative_expansion}). We consider a more generic Hamiltonian 
\begin{equation}
    \hat H(t) = \hat H_0(t) + \eta \hat H_1(t)\ .
    \label{eq:S_pert_H}
\end{equation}
We want to write a perturbative expansion for the unitary evolution operator $\hat U(t_0,t)$ under the assumption $\eta\ll1$:
\begin{equation}
    \hat U(t_0,t) = \hat U_0(t_0,t) + \eta \hat U_1(t_0,t) + \eta^2 \hat U_2(t_0,t) +...\ .
    \label{eq:S_pert_U}
\end{equation}
The operator $\hat U$ follows 
\begin{equation}
    \hat H(t)\hat U(t_0,t) \psi(t_0) = i\hbar \frac{\partial}{\partial t} \hat U(t_0,t) \psi(t_0)
    \label{eq:S_pert_shrodinger}
\end{equation}
and substituting $\hat H,\hat U$ from Eqs~(\ref{eq:S_pert_H},\ref{eq:S_pert_U}) for $\eta=0$ yields the equation defining the 0-th order term
\begin{equation}
    \hat H_0(t)\hat U_0(t_0,t) \psi(t_0) = i\hbar \frac{\partial}{\partial t} \hat U_0(t_0,t) \psi(t_0)
    \label{eq:S_pert_0th_eq}
\end{equation}
which we integrate to obtain 
\begin{equation}
    \hat U_0(t_0,t) = \exp\left[-\frac{i}{\hbar}\int_{t_0}^t\hat H_0(\tau)d\tau\right]\ .
\end{equation}
We now write the first-order equation by injecting $\hat H,\hat U$ from Eqs.~(\ref{eq:S_pert_H},\ref{eq:S_pert_U}) into Eq.~(\ref{eq:S_pert_shrodinger}), neglecting terms in $\eta^n,n>1$, and making use of Eq.~(\ref{eq:S_pert_0th_eq}) to obtain 
\begin{equation}
    \left(\hat H_1(t)\hat U_0(t_0,t)+\hat H_0(t)\hat U_1(t_0,t)\right) \psi(t_0) = i\hbar \frac{\partial}{\partial t} \hat U_1(t_0,t) \psi(t_0)\ ,
    \label{eq:S_pert_1st_eq}
\end{equation}
providing the first order contribution
\begin{equation}
    \hat U_1(t_0,t) = -\frac{i}{\hbar}\hat U_0(t_0,t)\int_{t_0}^t U_0(\tau,t_0)\hat H_1(\tau)\hat U_0(t_0,\tau)d\tau\ .
    \label{eq:S_pert_U1}
\end{equation}
Injecting this expression for $\hat U_1(t_0,t)$ in Eq~(\ref{eq:S_pert_1st_eq}), and utilizing the fact that $\hat U_0(t_0,t)\hat U_0(t,t_0)= \mathds{1}$ demonstrates the validity of Eq.~(\ref{eq:S_pert_U1}).

\subsection{Sensitivity}
\label{sec:sensitivity}
The sensitivity of this technique is given by Eq.~(\ref{eq:S_sensitivity_1}).
It is maximum at the highest possible value for $A_y$.
For pulses using $\sin^2$ ramping, this is achieved by maximizing the ramp duration
\begin{equation}
    a_{2\pi}(t) = \sin^2(\pi t/T),\ 
\end{equation}
yielding $A_y\approx1$ and a simple formula for the sensitivity
\begin{equation}
    P_{\ket{0}} \simeq 1-\frac14\left(N\phi'\right)^2\ .
\end{equation}
The characterization of $\phi'$ can be disturbed by factors split into three categories.
First, constant offsets in the signal amplitude or qubit frequency, which preserve the coherence of the pulse train.
These will decrease the contrast $P_{\ket{0}}$ following the same quadratic scaling $\propto N^2$ as phase variations, and therefore drifts in these parameters should remain small.
As an example of what ``small'' means in this context, the case of amplitude drifts is captured by Eq.~(\ref{eq:S_P0}) through the parameter $\epsilon$, and we see that any amplitude drift should satisfy $|\epsilon A_x|\ll|\phi' A_y|$.
Second are decoherence effects characterized by the time $T_2$, which reduce the contrast by the same amount for each pulse, reducing the sensitivity following
\begin{equation}
    P_{\ket{0}} \simeq 1-N^2\frac{\phi'^2}{4}-N\frac{T}{4T_2}\ ,
\end{equation}
when $NT \ll T_2$.
To measure $\phi'$, coherence should be preserved over the pulse train $NT \ll T_2$, whilst using a sufficiently large number of pulses that the consequence of phase variations dominates that of decoherence $N^2\phi'^2\gg NT/T_2$.
Combining these two conditions, we reach the requirement $\phi'\gg T/T_2$ (satisfied by three orders of magnitude in this experiment).
Finally, state-preparation and measurement errors will also affect $P_{\ket{0}}$, but lead to a loss of contrast independent of the number of pulses that can be compensated by an increased number of pulses.

\clearpage
\onecolumngrid
\section{Supporting measurements}
\label{sec:supporting_measurements}

\begin{figure*}[h!]
    \centering
    \includegraphics[width=0.9\textwidth]{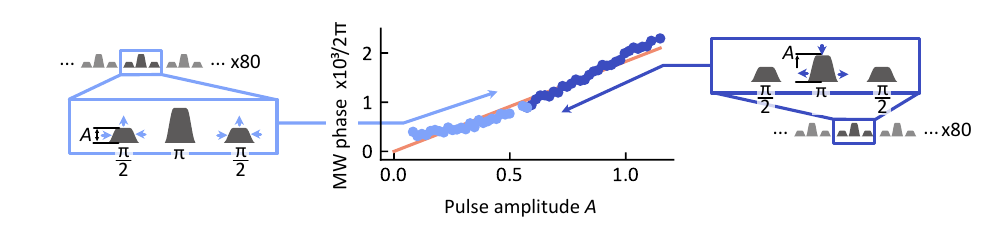}
    \caption{
        \textbf{Alternative measurement of the phase-amplitude relation.}
        A train of 80 $\pi/2-\pi-\pi/2$ pulse sequences is delivered to the qubit.
        In the absence of phase variations between the pulses, the sequence should have no net effect on the qubit.
        If a phase-amplitude relation exists however, maintaining a large difference in amplitude between the $\pi/2$ and $\pi$ pulses leads to phase differences between these two pulses.
        The pulse sequence then induces a change in the qubit state, which coherently adds up with the number of sequences, from which the phase difference can be inferred.\\
        We vary the amplitude and duration of the $\pi$ or $\pi/2$ pulses, maintaining an approximately constant pulse area, to map out the phase-amplitude relationship.
        The measurement has to be carried out in two steps however, if the full range of microwave amplitudes is to be explored.
        Indeed, the sensitivity of this measurement increases with the difference in amplitude between the pulses.
        In the light blue data, we start with the shortest, highest amplitude $\pi$ pulse possible, and increase the $\pi/2$ amplitude until phase variations are difficult to resolve.
        In the dark blue data, we start with $A\sim 0$ amplitude $\pi/2$ pulse, and decrease the amplitude of $\pi$ pulses until sensitivity is lost.\\
        Maintaining the correct pulse area is complicated by the lack of control over pulse time (relative to the high degree of control over amplitude), as well as amplitude non-linearity (the amplitude requested from the AWG is not proportional to the field amplitude at the ion).
        In practice, we find that a slight adjustment of the amplitude of one of the pulses is required for each point shown in this figure.
        Each measurement (light/dark blue points) are thus extracted from different 2-dimensional parameter scans, which are considerably longer to carry out than the calibration demonstrated in Fig.~\ref{fig:fig3}(a).
        This measurement does, however, demonstrate an approximately linear phase-amplitude relationship, confirming the findings reported in the main text.
    }
    \label{fig:nonlinearity_ion}
\end{figure*}

\begin{figure*}[h]
    \centering
    \includegraphics[width=0.9\textwidth]{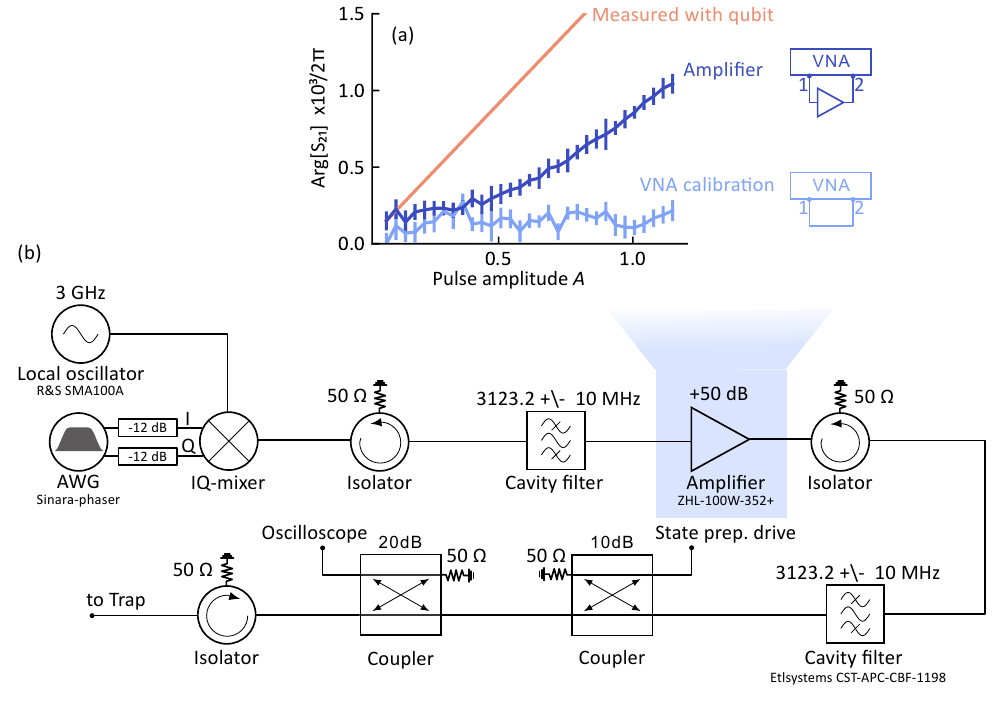}
    \caption{\textbf{Nonlinearity in the microwave amplifier} (a) The measured phase shift induced by the amplifier (Arg[$S_\text{21}$]) used in the qubit microwave drive chain is shown in dark blue.
    The measurement is carried out using a vector network analyzer (Rohde\&Schwarz ZVL) with a precision of $~2\pi 10^{-4}$ radians, measured in the control experiment shown in light blue where the amplifier is removed from the measured network.
    The amplifier contributes to $\phi'= 2\pi\times 0.8\times10^{-3}$ of the total phase shift $\phi'= 2\pi\times 1.8\times10^{-3}$ measured using the qubit (orange).
    (b) Microwave setup used to drive the qubit, showing the multitude of potentially nonlinear components which could produce phase shifts.
    All the components downstream from the IQ mixer were measured, collectively and individually, but only the amplifier showed a phase variation exceeding the precision of the VNA measurement, and showed a statistically identical phase non-linearity when measured in series with the other components.
    VNA measurements are limited here because (1) the AWG itself may introduce some non-linearity, for example if the relative amplitude of the I/Q channels varies, and this cannot be captured by a VNA measurement, and (2) after the AWG, the signal is up-converted using an IQ mixer which we are unable to fully characterize with our VNA which only features a single source, and a same-frequency receiver.
    One channel of the IQ mixer can be measured with the VNA using a second mixer for down-conversion, however no significant phase variations were observed.
    We attempt to circumvent the limitations of this measurement with the additional measurement of Fig.~\ref{fig:nonlinearity_scope}.
    }
    \label{fig:nonlinearity_amplifier}
\end{figure*}

\begin{figure*}[h]
    \centering
    \includegraphics[width=0.9\textwidth]{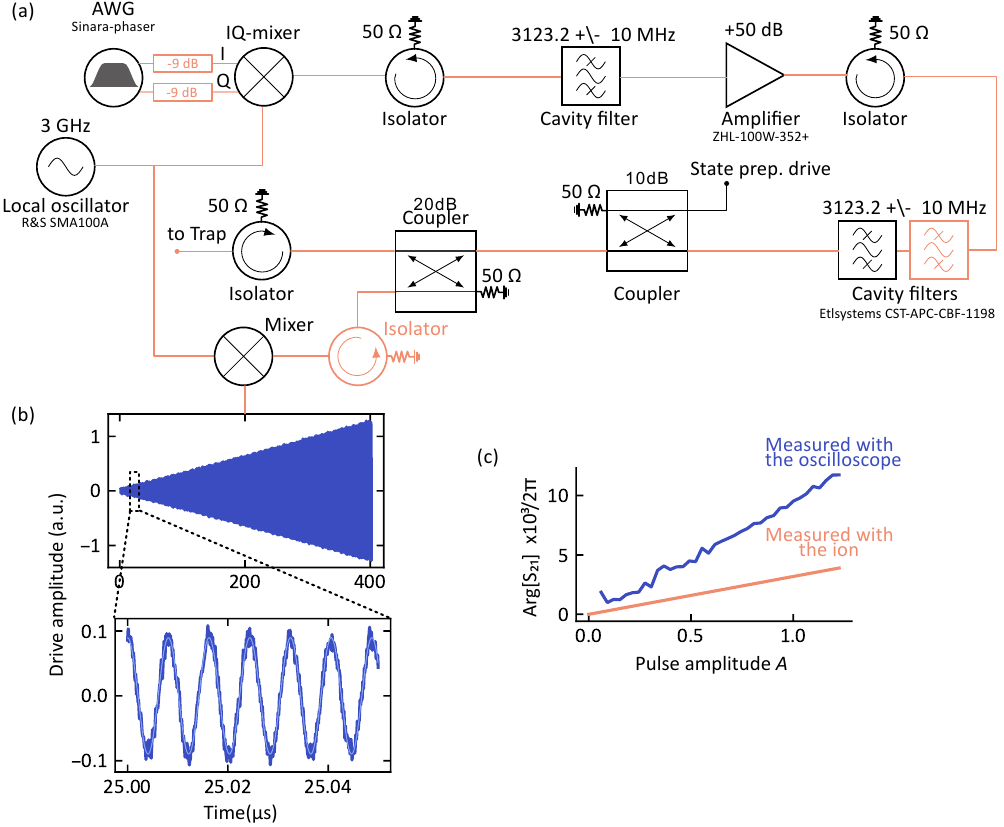}
    \caption{\textbf{Nonlinearity in the microwave chain} 
    Here we present an alternative ex-situ characterization of phase variations.
    (a) The experiment is carried out at a later date, after substantial modifications to the microwave chain highlighted in orange.
    Notably the cabling and attenuations have been modified, changing the microwave power in the nonlinear elements of the signal chain.
    As a result, in this setup we measure a larger $\phi'= 2\pi\times 3.2\times10^{-3}$.
    (b) We ramp the microwave amplitude in a series of steps, and a down-converted signal is captured by an oscilloscope.
    For each step of the signal, the captured sinusoid is fitted to extract the signal phase, as shown in the lower panel.
    (c) The phase variation with amplitude is then compared to the characterization performed with the ion.
    As with the VNA measurement shown in Fig.~\ref{fig:nonlinearity_amplifier}, the phase variation is of the correct order of magnitude, but fails to faithfully capture the behavior at the ion.
    This again shows the value of an in-situ characterization.
    }
    \label{fig:nonlinearity_scope}
\end{figure*}

\end{document}